\documentclass[12pt]{article}

\usepackage{amsmath}
\usepackage{amssymb}
\usepackage[dvips]{graphicx}

\usepackage{cite}

\makeatletter
 
  \@addtoreset{equation}{section}
 \makeatother

\begin{document}
\setlength{\oddsidemargin}{0cm}
\setlength{\baselineskip}{7mm}

\begin{titlepage}

~~\\

\vspace*{0cm}
    \begin{Large}
       \begin{center}
         {On the Structure Constants of Volume Preserving Diffeomorphism Algebra}
       \end{center}
    \end{Large}
\vspace{1cm}

\begin{center}
           Matsuo S{\sc ato}\footnote
           {
e-mail address : msato@cc.hirosaki-u.ac.jp}\\
      \vspace{1cm}
       
         {\it Department of Natural Science, Faculty of Education, Hirosaki University\\ 
 Bunkyo-cho 1, Hirosaki, Aomori 036-8560, Japan}

\end{center}

\hspace{5cm}

\begin{abstract}
\noindent
Regularizing volume preserving diffeomorphism (VPD) is equivalent to a long standing problem, namely regularizing Nambu-Poisson bracket. In this paper, as a first step to regularizing VPD, we find general complete independent basis of VPD algebra. Especially, we find complete independent basis that give simple structure constants, where three area preserving diffeomorphism (APD) algebras are manifest. This implies that an algebra that regularizes VPD algebra should include three $u(N)$ Lie algebras.

\end{abstract}

\vfill
\end{titlepage}
\vfil\eject

\setcounter{footnote}{0}

\section{Introduction}
\setcounter{equation}{0}
Area preserving diffeomorphism (APD) algebra is regularized by $u(N)$ Lie algebra. Actually, a large N limit of structure constants of $u(N)$ Lie algebra in the 't Hooft basis reduces to those of APD algebra defined on $T^2$\cite{Fairlie:1988qd, Hoppe:1988gk, Fairlie:1989vv, deWit:1989vb}. Because APD algebra is generated by Poisson bracket, it is regularized by Lie bracket of $u(N)$ Lie algebra. This structure induces that the Heisenberg picture of quantum mechanics reduces to the canonical formalism of classical mechanics in the classical limit. Another application is that one can show that  BFSS matrix theory and IIB matrix model contain the lightcone supermembrane and the type IIB superstring, respectively by using this regularization \cite{Banks:1996vh, deWit:1988ig, Ishibashi:1996xs}.

On the other hand, regularizing Nambu-Poison bracket is a long standing problem\footnote{For example, if the problem is solved, one should be possible to show that a three algebra model of M-theory \cite{Sato:2009mf, Sato:2010ca, Sato:2011gi, Sato:2013mja, Sato:2013iqa, Sato:2013aoa} contains the semi-lightcone supermembrane.}\cite{Nambu:1973qe, Floratos:1989au, Chatterjee:1995iq, Hoppe:1996xp, Dito:1996xr, 1998AIPC..453..107S, Awata:1999dz, Minic:1999js, Minic:2000kw, Hofman:2001zt, Curtright:2002fd, Shimada:2003ks, Zachos:2003as, DeCastro:2003jc, Axenides:2006ms, Chu:2008qv, Axenides:2008rn, Akerblom:2009gx, Axenides:2009kr, Sasakura:2010rb, Gustavsson:2010nc, 2011arXiv1110.0134B, Sato:2012ti, Jurco:2012yv}. As in the case of APD, Nambu-Poisson bracket generates volume preserving diffeomorphism (VPD) algebra. In this paper, as a first step to regularizing Nambu-Poisson bracket, we search for several independent basis of VPD algebra and obtain simple structure constants.

\vspace{1cm}

\section{General Complete Independent Basis of VPD Algebra}
VPD is a diffeomorphism $x^i \to y^i(x)$ ($i=1,2,3$) that satisfies $\mbox{det}\partial_i y^j(x)=1$. Then, the infinitesimal transformation $y^i(x)=x^i+ \delta x^i(x)$ satisfies 
\begin{equation}
\partial_i \delta x^i(x)=0. \label{VPDequation}
\end{equation}
$\delta x^i(x) = \epsilon^{i j k}\partial_j f(x) \partial_k g(x)$ satisfy this equation. Transformations of a scalar field generated by these solutions are given by \begin{eqnarray}
\delta Z(x) &\equiv& \delta x^i(x) \partial_i Z(x) \nonumber \\
&=& \epsilon^{i j k} \partial_i f(x) \partial_j g(x) \partial_k Z(x) \nonumber \\
&=& \{ f(x), g(x), Z(x)\}.
\end{eqnarray}
This implies that Nambu-Poisson bracket generates VPD. The transformations 
\begin{equation}
\delta=\delta x^i(x) \partial_i=\epsilon^{i j k} \partial_i f(x) \partial_j g(x) \partial_k \label{VPDcontgeneratror}
\end{equation}
form VPD algebra.

APD is a two-dimensional analogue of VPD. Infinitesimal transformations 
\begin{equation}
\delta=\delta X^I(Y) \partial_I=\epsilon^{I J} \partial_I F(Y)\partial_J \label{APDcontgenerator}
\end{equation}
 on $T^2$ $(I,J=1,2)$,
where
\begin{equation}
\partial_I \delta X^I(X)=0, \label{APDequation}
\end{equation}
 are spanned by generators
\begin{equation}
\delta(A)=ie^{iAY}\epsilon^{I J} A_I \partial_J,
\label{APDgenerator}
\end{equation}
which are obtained by substituting $ F(Y)=e^{iAY}$ to (\ref{APDcontgenerator}).

On the other hand, complete independent basis of VPD cannot be obtained by substituting $ f(x)=e^{iax}$ and $g(x)=e^{ibx}$ to (\ref{VPDcontgeneratror}) because $\delta x^i(x)$ is a local vector in three dimensions. We need to solve (\ref{VPDequation}). In the case of APD, (\ref{APDgenerator}) are complete independent solutions of (\ref{APDequation}). On $T^3$, we make a Fourier transformation,
$\delta x^i(x)= \sum_{a} v^i(a)e^{iax}.$ (\ref{VPDequation}) implies 
\begin{equation}
a_i v^i(a)=0. \label{orthogonal}
\end{equation}

An independent solution of (\ref{orthogonal}) is given by 
\begin{eqnarray}
\bar{v}_1&=&(-a_2, a_1, 0) \nonumber \\
\bar{v}_2&=&a \times \bar{v}_1=(-a_1 a_3, -a_2 a_3, a_1^2+a_2^2), \label{complicatedv1}
\end{eqnarray}
for $a=(a_1, a_2, a_3)$ (except $a_1=a_2=0$), and
\begin{eqnarray}
\!\!\!\!\!\!\!\!\! \!\!\!\!\!\!\!\!\!\!\!\!\!\!\!\!\!\!\!\!\!\!\!\!\!\!\!\!\!\!\!\!\!\!\!\!\!\!\!\!\!\!\!\!\!\!\!\!\!\!\!\!\!\!\bar{v}_1^{'}&=&(1,0,0) \nonumber \\
\!\!\!\!\!\!\!\!\! \!\!\!\!\!\!\!\!\!\!\!\!\!\!\!\!\!\!\!\!\!\!\!\!\!\!\!\!\!\!\!\!\!\!\!\!\!\!\!\!\!\!\!\!\!\!\!\!\!\!\!\!\!\!\bar{v}_2^{'}&=&(0,1,0), \label{complicatedv2}
\end{eqnarray}
for $a=(0,0,a_3)$.

The corresponding VPD generators are given by 
\begin{eqnarray}
S_1(a)&=&e^{iax}\bar{v}_1^i \partial_i= e^{iax}(-a_2 \partial_1 + a_1 \partial_2) \nonumber \\
S_2(a)&=&e^{iax}\bar{v}_2^i \partial_i= e^{iax}(-a_1 a_3 \partial_1 - a_2 a_3 \partial_2 + (a_1^2+a_2^2) \partial_3) \label{complicatedbasis1} \\
S_1'(0,0,a_3)&=&e^{ia_3x^3}\bar{v}_1^{'i} \partial_i= e^{ia_3x^3}\partial_1
\nonumber \\
S_2'(0,0,a_3)&=&e^{ia_3x^3}\bar{v}_2^{'i} \partial_i= e^{ia_3x^3}\partial_2,
\label{complicatedbasis2}
\end{eqnarray}
which form VPD algebra 
\begin{eqnarray}
&&[S_1(a), S_1(b)]=i (a_1 b_2 - a_2 b_1) S_1(a+b) \nonumber \\
&&[S_2(a), S_2(b)]=i \alpha S_1(a+b)+ i \beta S_2(a+b) \nonumber \\
&&[S_1(a), S_2(b)]=i \gamma S_1(a+b)+ i \delta S_2(a+b) \nonumber \\ 
&&[S_1(a), S_1'(0,0,b_3)]=-i a_1 S_1(a+b) \nonumber \\
&&[S_2(a), S_1'(0,0,b_3)]=-i a_2 b_3 S_1(a+b)-ia_1 S_2(a+b) \nonumber \\
&&[S_1(a), S_2'(0,0,b_3)]=-i a_2 S_1(a+b) \nonumber \\
&&[S_2(a), S_2'(0,0,b_3)]=i a_1 b_3 S_1(a+b)-ia_2 S_2(a+b) \nonumber \\
&& [S_1'(0,0,a_3), S_1'(0,0,b_3)]=[S_2'(0,0,a_3), S_2'(0,0,b_3)]=[S_1'(0,0,a_3), S_2'(0,0,b_3)]=0, \nonumber \\
\end{eqnarray}
where
\begin{eqnarray}
\alpha&=&\frac{1}{(a_1+b_1)^2+(a_2+b_2)^2}(a_2 b_1-a_1 b_2)((a_1^2+a_2^2)b_3^2 + (b_1^2+b_2^2) a_3^2-2 a_3 b_3(a_1b_1+a_2 b_2))  
\nonumber \\
\beta&=& \frac{1}{(a_1+b_1)^2+(a_2+b_2)^2}((a_1^2+a_2^2)b_3((a_1+b_1)b_1+(a_2+b_2)b_2) \nonumber \\
&&\qquad\qquad\qquad\qquad\qquad - (b_1^2+b_2^2)a_3((a_1+b_1)a_1+(a_2+b_2)a_2))
\nonumber \\
\gamma&=& \frac{1}{a_1+b_1}(\frac{1}{ (a_1+b_1)^2+(a_2+b_2)^2}(b_1^2+b_2^2)(a_1 b_2-a_2 b_1)(a_2+b_2)(a_3+b_3) \nonumber \\
&&\qquad\qquad -b_2b_3(a_1b_2-a_2b_1)-a_1(-b_3(a_1b_1+a_2b_2)+a_3(b_1^2+b_2^2)))  
\nonumber \\
\delta&=& \frac{1}{(a_1+b_1)^2+(a_2+b_2)^2} (b_1^2+b_2^2)(a_1 b_2-a_2 b_1).
\end{eqnarray}
This algebra has a complicated form because the basis (\ref{complicatedbasis1}) and (\ref{complicatedbasis2}) are complicated.

General independent solutions of (\ref{orthogonal}) are given by 
\begin{eqnarray}
v_1^i&=&\epsilon^{ijk}a_j l_k(a) \nonumber \\
v_2^i&=&\epsilon^{ijk}a_j m_k(a), \label{v1v2}
\end{eqnarray}
where $a$, $l(a)$, $m(a)$ are all independent for all $a$. The corresponding generators are given by 
\begin{eqnarray}
T_1(a)&:=&e^{i a \cdot x} \mbox{det}(l a \partial) \nonumber \\
T_2(a)&:=&e^{i a \cdot x} \mbox{det}(m a \partial), \label{T1T2}
\end{eqnarray}
where $\mbox{det}(abc):= \epsilon^{ijk}a_i b_j c_k$. 

If we choose $l=(0,0,1)$ and $m=(-a_2, a_1,0)$ for $a=(a_1,a_2,a_3)$ (except $a_1=a_2=0$), (\ref{v1v2}) and (\ref{T1T2}) represent (\ref{complicatedv1}) and (\ref{complicatedbasis1}), respectively. If we choose $l=(0,-\frac{1}{a_3},0)$ and $m=(\frac{1}{a_3},0,0)$ for $a=(0,0,a_3)$, (\ref{v1v2}) and (\ref{T1T2}) represent (\ref{complicatedv2}) and (\ref{complicatedbasis2}), respectively.

\section{Simple Structure Constants of VPD Algebra}
In this section, we search for complete independent basis that give more simple structure constants. Although (\ref{T1T2}) for constant $l$ and $m$ are not independent in a part of the region of $a$ where $a$ is on a plane spanned  by $l$ and $m$, we can calculate commutation relations among (\ref{T1T2}) for constant $l$ and $m$, and then obtain simple relations
\begin{eqnarray}
&&[T_1(a), T_1(b)]=i \mbox{det}(lab) T_1(a+b) \\
&&[T_2(a), T_2(b)]=i \mbox{det}(mab) T_2(a+b) \\
&&[T_1(a), T_2(b)]=i \frac{1}{\mbox{det}(lm(a+b))}(\mbox{det}(mab)\mbox{det}(lma)T_1(a+b)
                                            +\mbox{det}(lab)\mbox{det}(lmb)T_2(a+b)). \nonumber \\
\end{eqnarray}

For example, if we choose $l=(0,0,1)$ and $m=(1,0,0)$, we obtain $v_1=(-a_2, a_1, 0)$, $v_2=(0, -a_3, a_2)$, and the corresponding generators
\begin{eqnarray}
U_1(a)&=&e^{iax}(-a_2 \partial_1 + a_1 \partial_2) \nonumber \\
U_2(a)&=&e^{iax}(- a_3 \partial_2 + a_2 \partial_3).
\end{eqnarray}
In this case, for $a_2=0$, $v_1=(0, a_1, 0)$ and $v_2=(0, -a_3, 0)$ are dependent, and thus $U_1(a)$ and $U_2(a)$ are. 

Then, we choose a step function, $m=(1,0,0)$ for $a_2 \neq 0$ and $m=(0,1,0)$ for $a_2 = 0$. When $m=(0,1,0)$ we have $v_3=(a_3, 0, -a_1)$ and 
\begin{equation}
U_3(a)=e^{iax}(a_3 \partial_1 - a_1 \partial_3),
\end{equation}
which is independent of $U_1(a)$ and $U_2(a)$ for $a_2 = 0$.
After considering $v_1=0$ when $a_1=a_2=0$, we have a complete set of independent generators 
\begin{eqnarray}
&&U_1(a) \mbox{  (except $a_1=a_2=0$)} \nonumber \\
&&U_2(a) \mbox{  (except $a_2=0$)} \nonumber \\
&&U_2(0,0,a_3) \nonumber \\
&&U_3(a_1, 0, a_3).
\end{eqnarray}
In fact, for each $a$ there are independent two generators;
\begin{eqnarray}
&&U_1(a) \mbox{  and  } U_2(a) \mbox{  for $a_2 \neq 0$} 
\nonumber \\
&&U_1(a) \mbox{  and  } U_3(a) \mbox{  for $a_2=0$ and $a_1 \neq 0$} 
\nonumber \\
&&U_2(a) \mbox{  and  } U_3(a) \mbox{  for $a_2=a_1=0$}.\end{eqnarray}
Then, we obtain simple structure constants of VPD algebra, 
\begin{eqnarray}
&&[U_1(a), U_1(b)]=i (a_1 b_2 - a_2 b_1) U_1(a+b) \label{APD1}\\
&&[U_2(a), U_2(b)]=i (a_2 b_3 - a_3 b_2) U_2(a+b) \label{APD2}\\
&&[U_3(a_1, 0, a_3), U_3(b_1, 0, b_3)]=i (a_3 b_1 - a_1 b_3) U_3(a_1+b_1, 0, a_3+b_3) \label{APD3}\\
&&[U_1(a), U_2(b)]=i \frac{1}{a_2 +b_2} (a_2(a_2 b_3 - a_3 b_2) U_1(a+b)
+b_2(a_1 b_2 - a_2 b_1) U_2(a+b)) \\
&&[U_1(a), U_3(b_1, 0, b_3) ]=i ((-a_1b_3+a_3b_1+b_1b_3)U_1(a+b)+b_1^2U_2(a+b))  \\
&&[U_1(a), U_3(b_1, 0, b_3)]=i (-b_3^2 U_1(a+b)+(a_3b_1-b_3a_1-b_1b_3)U_2(a+b)).
\end{eqnarray}
From (\ref{APD1}), (\ref{APD2}) and (\ref{APD3}), one can see three APD algebras corresponding to $(x^1, x^2)$, $(x^2, x^3)$ and $(x^3, x^1)$ planes.

\section{Conclusion and Discussion}
\setcounter{equation}{0}
In this paper,  we found general complete independent basis of VPD algebra. Especially, we found complete independent basis that give simple structure constants where three APD algebras are manifest. This implies that an algebra that regularizes VPD algebra should include three $u(N)$ Lie algebras.

\vspace*{1cm}

\section*{Acknowledgements}
We would like to thank T. Asakawa, K. Hashimoto, N. Ikeda, N. Kamiya, H. Kunitomo, T. Matsuo, S. Moriyama, K. Murakami, J. Nishimura, S. Sasa, P. Schupp, F. Sugino, T. Tada, S. Terashima, S. Watamura, K. Yoshida, and especially H. Kawai and A. Tsuchiya for valuable discussions. This work is supported in part by Grant-in-Aid for Young Scientists (B) No. 25800122 from JSPS.

\vspace*{0cm}

\bibliographystyle{unsrt}
\bibliography{VPDalg}

\end{document}